\documentclass[12pt]{article}
\usepackage{psfrag,epsfig,amsmath,amssymb}

\makeatletter
\renewcommand{\section}{\setcounter{equation}{0}\@startsection
  {section}%
  {1}%
  {0pt}%
  {-1\baselineskip}%
  {0.4\baselineskip}%
  {\bfseries\large}}%
\renewcommand{\subsection}{\@startsection
  {subsection}%
  {2}%
  {0pt}%
  {-0.75\baselineskip}%
  {0.2\baselineskip}%
  {\bfseries}}%
\renewcommand{\subsubsection}{\@startsection
  {subsubsection}%
  {3}%
  {0pt}%
  {-0.5\baselineskip}%
  {0.1\baselineskip}%
  {\sc}}%
\makeatother

\makeatletter
 \newcommand\figcaption{\def\@captype{figure}\caption}
\makeatother

\def\a{\alpha}
\def\b{\beta}

\def\ga{\gamma}

\def\la{\lambda}

\def\m{\mu}
\def\n{\nu}
\def\r{\rho}
\def\s{\sigma}

\def\th{\theta}

\def\Dirac{{D\mkern-12mu/}}

\def\prslash{{\partial\mkern-9mu/}}
\def\pslash{{p\mkern-8mu/}{\!}}
\def\kslash{{k\mkern-8mu/}{\!}}

\def\prslash{{\partial\mkern-9mu/}}    %%_standard_Dirac_operator

\def\pslash  {{p\mkern-7mu/}}

\def\bp{\text{\tiny{BPST}}}
\def\id{{\rm{I}\!\rm{I}}}

\def\id3x{\int\!\! d^3\!\vec{x}}
\def\idx{\int\!\! d^4\!x}
\def\iDx{\int\!\! d^D\!x}

\def\rig>{\right>}
\newcommand{\bea}{\begin{eqnarray}}
\newcommand{\eea}{\end{eqnarray}}
\newcommand{\beann}{\begin{eqnarray*}}
\newcommand{\eeann}{\end{eqnarray*}}
\newcommand{\ba}{\begin{array}}
\newcommand{\ea}{\end{array}}

\newcommand{\Tr}{\mathbf{Tr}}

\def\Psib{\bar{\Psi}}
\def\g5{\gamma_{5}}

\def\prslash {{\partial\mkern-9mu/}}  %operador Dirac
\def\pslash  {{p\mkern-7mu/}}

\def\idx3{\int\! d^{3}\!\vec{x}\,}
\def\idx{\int\! d^{4}\!x\,}

 \def\Psib{\bar{\Psi}}

 \def\Dirac{{D\mkern-12mu/}\,}

 \def\prslash {{\partial\mkern-9mu/}}  %operador Dirac
 \def\pslash  {{p\mkern-7mu/}}

\def\bk {\bar{k}} %%terminos hatted
 \def\bp {\bar{p}}
 \def\bq {\bar{q}} 
 \def\bg {\bar{\gamma}}
 \def\bdelta {\bar{\delta}}
 \def\Db {{\partial}_{\beta}}

 \def\ab {a_{\beta}}
 \def\am {a_{\mu}}

 \def\g {\gamma}
 \def\mi {{\mu_1}}

 \def\a {\alpha}
\def\b {\beta}
\def\r {\rho}
 \def\s {\sigma}

 \def\Tr{\text{Tr}}

\begin{document}
\begin{titlepage}

\hfill{NSF-KITP-09-178}\\
\rightline{FTI/UCM 101-2009}\vglue 10pt
\begin{center}

{\Large \bf Renormalisability of noncommutative GUT inspired field theories  with anomaly safe groups}\\
\vskip 0.2 true cm {\rm C.P. Mart\'{\i}n$^{\dagger,}$\footnote{E-mail: carmelo@elbereth.fis.ucm.es}
and  C. Tamarit$^{\dagger\dagger,}$}\footnote{E-mail: tamarit@kitp.ucsb.edu}
\vskip 1pt  $^\dagger${\it Departamento de F\'{\i}sica Te\'orica I,
Facultad de Ciencias F\'{\i}sicas\\
Universidad Complutense de Madrid,
 28040 Madrid, Spain}
 \vskip -1pt $^{\dagger\dagger}${\it Kavli Institute for Theoretical Physics, University of California\\
 Santa Barbara, CA, 93106-4030, USA}\\
\vskip -1cm

\end{center}

{\leftskip=50pt \rightskip=50pt \noindent We consider noncommutative GUT inspired field theories 
formulated within the enveloping-algebra formalism  for anomaly safe compact simple gauge groups.
Our theories have only gauge fields and fermions, and we compute the UV  divergent part of the one-loop  
background-field effective action involving two fermionic fields at first order 
in the noncommutativity parameter $\theta$. We show that, if the second-degree Casimir has the 
same value for all the irreps furnished by the fermionic multiplets of the model, then,  that UV  divergent part can be 
renormalised  by carrying out  multiplicative renormalisations of the coupling constant, $\theta$ and the fields,   
along with the inclusion of  $\theta$-dependent counterterms 
which vanish upon imposing the equations of motion. 
These $\theta$-dependent counterterms have no physical effect since they vanish on-shell. This result along with the 
vanishing of the UV divergent part of the fermionic four-point functions leads to the unexpected conclusion that 
the one-loop matter sector of the background-field effective action of these theories is one-loop multiplicatively
renormalisable on-shell.   We also show that the background-field effective action of the gauge sector 
of the theories considered here receives no $\theta$-dependent UV divergent contributions at one-loop. We thus
conclude that these theories are on-shell one-loop multiplicatively renormalisable at first order in $\theta$.

\par }

\vspace{-1pt} \noindent
{\em PACS:} 11.10.Gh, 11.10.Nx, 11.15.-q, 12.10.-g.\\
{\em Keywords:} Renormalization, Regularization and Renormalons, Non-commutative geometry. 
\end{titlepage}

%----------------------------------------------------- Paper

\setcounter{page}{2}
\section{Introduction}

	Noncommutative gauge theories with simple groups can only be formulated with the so called enveloping algebra approach, which makes use of Seiberg-Witten maps to relate noncommutative gauge orbits to ordinary ones  \cite{Jurco:2001rq}. Since the Seiberg-Witten maps are generically obtained perturbatively in the noncommutativity parameters $\theta$, the resulting theories, which are invariant under ordinary gauge transformations, involve interaction terms at all orders in $\theta$. This, and the fact that $\theta$ has negative mass dimensions, seems to suggest that the theories are only meaningful as effective theories. However, some intriguing results seem to point towards a perturbative self-consistency of the theories: it could well be that for some models the structure imposed by the Seiberg-Witten maps survives quantum corrections, so that the divergences can be absorbed by both multiplicative renormalisations and by  physically irrelevant counterterms (e.g., couterterms which vanish on-shell). The first one of these results concerns the fact that the gauge anomaly cancellation conditions have been shown to be, to all orders in $\theta$, equal to their commutative counterparts \cite{Brandt:2003fx}; this allowed to formulate noncommutative extensions of the Standard Model \cite{Calmet:2001na}, and GUT theories \cite{Aschieri:2002mc}. Other results concern the renormalisability of the gauge sector at one-loop, observed for a variety of models independently of the matter content \cite{Wulkenhaar:2001sq,Buric:2002gm,Buric:2004ms,Buric:2005xe,Latas:2007eu,Buric:2006wm,Martin:2006gw,Martin:2009mu}; in fact, the matter determinants contributing to the one-loop gauge effective action are known to yield renormalisable contributions to all orders in $\theta$ , at least for non-chiral theories \cite{Martin:2007wv}.

	 Despite these auspicious results, the matter sector  --matter in the fundamental representation-- of the theories studied so far --having U(1) and SU(2) as gauge groups-- is nonrenormalisable \cite{Wulkenhaar:2001sq,Buric:2004ms, Martin:2006gw} and the lack of renormalisability can be traced back to  problematic divergences in four point functions of the matter fields . There are, however, promising exceptions: on the one hand, supersymmetric (S)U(N) theories with adjoint Majorana fermions in a vector multiplet have been shown to be one-loop renormalisable \cite{Martin:2007wv}, and, on the other, noncommutative GUT inspired theories with arbitrary groups and representations have been shown to be free of the unwelcomed four fermion divergences just mentioned  \cite{Buric:2007ix,Martin:2009sg}.
	
	In this paper we continue the study of the renormalisability of noncommutative GUT inspired theories with no scalar fields, by computing the UV divergent part of the effective action involving two fermion fields. Here, we do it for theories with anomaly safe compact simple gauge groups --groups for which the anomaly coefficient vanishes in all representations-- since among these groups one finds the
phenomenologically promising SO(10) and $\rm{E}_6$. These models  have, as a consequence of the anomaly cancellation condition, no  vertices of order one in $\theta$ in the bosonic part of the classical action, and hence are not sensitive to ambiguities in the trace over bosonic fields \cite{Aschieri:2002mc}. We use the background field method in the Feynman-background-field gauge in conjunction with 
dimensional regularisation  to reconstruct, at first order in  $\theta$, the full one-loop UV divergent contribution involving two fermion fields. We do so by using gauge invariance and working out  the pole part of the two- and three-point Green functions involving, respectively, two fermion fields and one gauge field and two fermion fields. The result is the following: whenever all the irreducible representations carried by the  fermion multiplets of the theory  share the same second-degree  Casimir, the UV divergences can be renormalised  by using multiplicative renormalisation of the coupling constant, the noncommutative matrix parameter $\theta^{\m\n}$ and the fields,  and by adding  $\theta$-dependent counterterms which vanish on-shell, i.e., upon imposing the equation of motion. These $\theta$-dependent counterterms which vanish on-shell have, of course, no physical effect.  
If one combines this result with  the absence of UV divergent contributions to the fermionic four-point function of these theories --see ref.~\cite{Martin:2009sg}--, one concludes that the one-loop matter sector of the theory  is renormalisable on-shell at first order in $\theta$; this is the first time that such property is shown to hold in a noncommutative theory with nonmajorana fermions. The requirement of a common second-degree Casimir for the matter representations can be fulfilled by using a single irreducible representation --as is commonly done in ordinary GUTs such as SO(10) and ${\rm E}_6$--, though our renormalisability result is valid for any choice of representation-- or combining a representation with its conjugate. Finally, once the matter sector has been seen to be renormalisable, we show by using formal arguments that there are no UV divergent contributions to the gauge sector which are of order one in $\theta$. We thus put forward, for the first 
time in the literature,  a huge family of noncommutative theories with chiral fermionic matter and GUT gauge groups which are one-loop 
renormalisable at first order in $\theta$, in the physical sense that only  the ordinary renormalisation of the coupling constant and a
new multiplicative renormalisation of the noncommutative matrix parameter $\theta^{\m\n}$ 
are needed to workout UV finite S-matrix elements: the counterterms --in particular, the a priori problematic $\theta$-dependent counterterms-- which are not given by the renormalisations of $\th$ and the coupling constant vanish on-shell. Recall that the free parameters of our classical noncommutative field theories are the coupling constant and $\theta^{\m\n}$. 
	
Now, to make sure that the Hamiltonian formulation of our theories is the elementary one --only one canonical momenta per generalised coordinate--,  we shall 
choose a  noncommutative matrix parameter $\theta^{\m\n}$ such that 
$\theta^{0 i}=0$, $i=1,2,3$. Hence, without lose of generality one can
say that  $\theta^{\m\n}$ is characterized by a single noncommutative parameter, say, $\theta$.

	The paper is organised as follows.  The theory is defined in section 2, where the computation by means of the background field method is also outlined. Section 3 includes the results of the computations of the UV divergent part of the effective action involving two fermion fields, whose renormalisability is discussed in section 4. Section 5 is dedicated to argue in favour of the renormalisability of the gauge sector. Conclusions are presented in section 6. We also include two appendices: appendix A provides the results for the divergent contributions to the Feynman diagrams involved in the computations of section 3, while appendix B gives the results for the beta  functions of the physical parameters of the theory, $g$ and $\theta$.
	
%%%%%%%%%%%%%%%%%%%%%%%%%%%%%%%%%%%%%%%%%%%%%%%%%%%%%%%
%%%%%%%%%%%%%%%%%%%%%%%%%%%%%%%%%%%%%%%%%%%%%%%%%%%%%%%
%%%%%%%%%%%%%%%%%%%%%%%%%%%%%%%%%%%%%%%%%%%%%%%%%%%%%%%

	\section{The theory and  their background field method quantisation}
	
	We shall consider a  general four-dimensional noncommutative GUT inspired theory with an arbitrary anomaly 
safe \cite{Brandt:2003fx}   compact simple gauge group and no scalar fields as formulated in ref.~\cite{Aschieri:2002mc}. We thus define the theory by means of a noncommutative left-handed chiral multiplet $\Psi$ in an arbitrary representation $\rho_\Psi$ of the gauge group, and an enveloping-algebra valued gauge field $A_\m$ with action
\begin{align}\label{S}
&{S}=\idx-\frac{1}{2g^2}\Tr F_{\m\n}\star F^{\m\n}+\Psib_L i\Dirac \Psi_L,\\
%%%
&\nonumber F_{\m\n}=\partial_\m A_\n-\partial_\n A_\m-i[A_\m, A_\n]_\star,\quad D_{\m}\psi_L=\partial_\m\Psi_L-i 
\rho_{\Psi}(A_\m)\star\Psi_L,
\end{align}
where, at first order in $\theta$, the noncommutative fields are defined  in terms of the ordinary ones 
$a_\m,\psi$ by the following standard Seiberg-Witten maps,
\begin{align}
\nonumber A_\mu&=a_\m+\frac{1}{4}\th^{\a\b}\{\partial_\a \am+f_{\a\m},\ab\}+O(\th^2),\\
%%%
\label{SW}
\Psi_L&=\psi_L-\frac{1}{2}\th^{\a\b}\rho_\psi(a_\a)\Db\psi_L+
\frac{i}{4}\th^{\a\b}\rho_\psi(a_\a) \rho_\psi(a_\b)\psi_L+O(\th^2).
\end{align}
Note that $\rho_{\psi}$  denotes an arbitrary unitary representation, which can be expressed as a direct sum of irreducible representations, $\rho_{\psi}=\bigoplus_{r=1}^F \rho^{r}_{\psi}$.  Accordingly, the fermion fields can be expressed as a direct sum of irreducible multiplets, $\Psi_{L}=\bigoplus_{r=1}^F \Psi_{L}^{r}$, $\psi_{L}=\bigoplus_{r=1}^F \psi_{L}^{r}$. 

Upon substituting eq.~\eqref{SW} in eq.~\eqref{S} and, then, expanding up to first order in $\theta$, one
obtains a classical action for the ordinary fields $a^a_\m$ and $\psi_L$. Within the enveloping-algebra formalism, 
the quantisation of the theory defined by this classical action defines  the corresponding  noncommutative field theory 
at first order in $\theta$. It has been shown in ref.~\cite{Aschieri:2002mc} that for compact simple gauge groups  
the anomaly cancellation  condition~\cite{Brandt:2003fx} makes the first order in $\theta$ contribution coming 
from the noncommutative Yang-Mills action in eq.~\eqref{S} vanish. So for the family of theories studied in this paper,
and at first order in $\theta$, the only classical noncommutative corrections  to the ordinary classical action come from
the fermionic action in eq.~\eqref{S}.

Since  we shall  formulate the Feynman rules of our theory in terms of 
ordinary Dirac fermions, we include in it  a  spectator right-handed fermion, 
as done in ref.~\cite{Martin:2009sg}
	\begin{equation*}
S\rightarrow S'= S+\idx\bar{\tilde\psi}_R i\prslash\tilde\psi_R,	\quad \psi=\left[\begin{array}{l}\tilde\psi_R\\
\psi_L\end{array}\right].%\label{Sshift}
\end{equation*}

Again as in ref.~\cite{Martin:2009sg}, we shall regularise the theory by means of dimensional regularisation in $D=4+2\epsilon$ dimensions,  using the BMHV scheme for defining $\gamma_5$  \cite{'tHooft:1972fi,Breitenlohner:1977hr}. In this scheme there is an infinity of  dimensionally regularised actions which reduce to \eqref{S} in the limit $D=4$, and which differ
from one another by evanescent operators \cite{Martin:1999cc}. Following \cite{Martin:1999cc} we will keep all the vector indices in interaction vertices ``four-dimensional'', i.e., contracted with the ``barred'' metric $\bar{g}_{\mu\nu}$;  we shall also define the dimensionally regularised $\theta^{\mu\nu}$ as being ``four-dimensional''. Furthermore, in our computations we shall discard any contribution which has a pole in $\epsilon$ but whose numerator is an evanescent operator. Since we shall be dealing with an anomaly free theory, these contributions involving evanescent operators have no physical effects at the one-loop level \cite{Martin:1999cc,Chanowitz:1979zu}--although they are needed at two loops and beyond \cite{Martin:1993yx} -- and are mere artifacts of the regularisation procedure. It is not difficult to convince onself that the famous one-loop 
$log\, (-Q^2/\mu^2)$ contributions to Green functions are uniquely fixed by the pole contributions to the effective action with no evanescent operators.

Our aim is to compute the one-loop UV divergent part of the effective action involving two fermion fields and no evanescent operator in a manifestly covariant approach, which allows to reconstruct the full contribution to the effective action from a minimum number of diagrams, as was done in ref.~\cite{Martin:2009mu}.  For this we use the background field method  \cite{Abbott:1980hw}. This method amounts to split the gauge field $a_\m$ in a background part $b_\m$ and a quantum part $q_\m,$
\begin{align}
\label{splitting}
a_\mu=b_\m+q_\m,
\end{align}
and choose a gauge fixing which preserves background gauge transformations
$$\delta q_\m=-i[q_\m,c],\,\delta b_\m=D[b]_\m c,\,\,D[b]_\m=\partial_\m-i[b_\m,\,].$$
This gauge fixing is 
\begin{align*}%\label{Sgf}
S_{gf}=-\frac{1}{2\alpha}\idx(D^{[b]}_\m q^\m)^2,\quad S_{gh}=\idx \bar cD^{[b]}_\m D^{[b+q]\m} c.
\end{align*}	
Adapting to our case the discussions in refs.~\cite{Abbott:1980hw} and \cite{Martin:2009mu}, introducing the classical fields $\hat b_\m,\hat\psi$, the 1PI functional is given by
\begin{align}
\Gamma[\hat b_\m,\hat \psi,\hat{\bar\psi}]=\idx\sum_k\sum_n \frac{-i}{ (k!)^2}\tilde\Gamma^{(n,k)}_{\scriptsize\begin{array}{l}
i_1,..,i_k; \,\, j_1,..,j_k;\begin{array}{l}\mi,..,\mu_n\\
a_1,..,a_k\end{array}
\end{array}}\prod_{l=1}^k\hat{\bar\psi}_{i_l}\prod_{p=1}^k\hat\psi_{j_p}\prod_{m=1}^n\hat b_{\m_m}^{a_m} .
\label{Gammaexp}
\end{align}
The previous effective action is gauge invariant under gauge transformations of the classical fields 
$\hat b_\m,\hat\psi,\hat{\bar\psi}$. The dimensionally regularised version of the effective action above is not strictly speaking gauge
invariant, i.e, it is gauge invariant modulo an evanescent operator which as we have argued above can be dropped for anomaly-
free theories in  UV divergent one-loop computations.

\begin{figure}[h]
\psfrag{p}{$p$}
\psfrag{m A}{$\mu, a$}
\psfrag{n B}{$\nu, b$}
\psfrag{r C}{$\r, c$}
\psfrag{m C}{$\m, a$}
\psfrag{i A}{$i s$}
\psfrag{j B}{$j t$}
\psfrag{q}{$q$}\psfrag{k}{$k$}
\psfrag{i s}{$is$}\psfrag{j t}{$j t$}\psfrag{k,a,m}{$a,\mu$}
\psfrag{k1}{$k_1$}\psfrag{k2}{$k_2$}\psfrag{k3}{$k_3$}
\psfrag{m1}{$a,\mu$}\psfrag{m2}{\hskip0.2cm$b,\nu$}
\begin{minipage}{0.27\textwidth}
%\begin{center}
\includegraphics[scale=1]{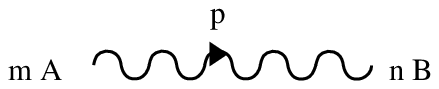}
%\end{center}
\end{minipage}%
\begin{minipage}{0.23\textwidth}
\begin{align*}
\leftrightarrow\frac{-ig^2 \delta^{ab}\eta^{\m\n}}{p^2+i\epsilon}
\end{align*}
\end{minipage}%
%%%%%%%%%%%%%%%%%%%%
\begin{minipage}{0.3\textwidth}
%\begin{center}
\includegraphics[scale=0.7]{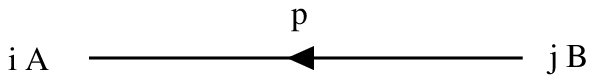}
%\end{center}
\end{minipage}%
\begin{minipage}{0.16\textwidth}
\begin{align*}
\leftrightarrow\frac{i (\pslash)_{ij}\delta_{st}}{p^2+i\epsilon}
\end{align*}
\end{minipage}\\
%%%%%%%%%%%%%%%%%%%%%%%%%%%%%%
\begin{minipage}{0.4\textwidth}
\flushleft\includegraphics[scale=0.7]{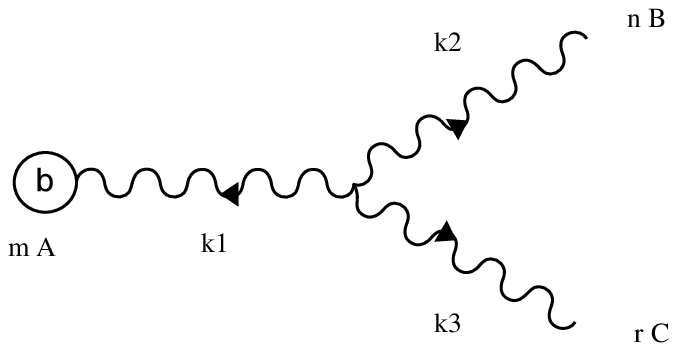}
\end{minipage}%
\begin{minipage}{0.55\textwidth}\flushleft
\begin{align*}
\leftrightarrow&\frac{1}{g^2}f^{abc}[\bar g^{\m\r}(\bar k_1-\bar k_3-\bar k_2)^\n+\bar g^{\n\r}(\bar k_3-\bar k_2)^\m\\
%%%%%%%%%%%%%%%%%%%
&+\bar g^{\m\n}(\bar k_2-\bar k_1+\bar k_3)^\r]
\end{align*}
\end{minipage}\\
%%%%%%%%%%%%%%%%%%%%%%%%%%
%%%%%%%%%%%%%%%%%%%%%%%%%%
%%%%%%%%%%%%%%%%%%%%%%%%%%
\begin{minipage}{0.22\textwidth}
\flushleft
\includegraphics[scale=0.55]{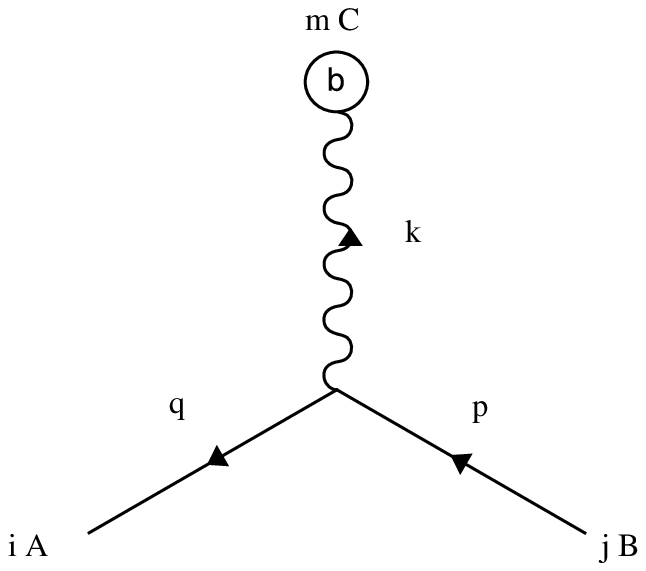}
\end{minipage}%
\begin{minipage}{0.18\textwidth}
\begin{align*}
&\leftrightarrow i(\bar\ga^\m)_{ij}(T^a)_{st}
\end{align*}
\end{minipage}
%%%%%%%%%%%%%%%%%
\begin{minipage}{0.22\textwidth}
\flushleft
\includegraphics[scale=0.55]{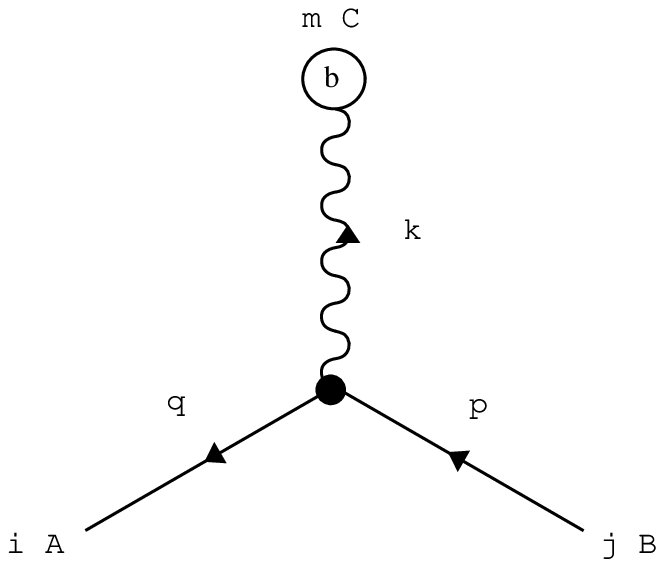}%
\end{minipage}%
%%%%%%%%%%%%%%%%%%%%%%%%%%%%%%%%%%%%%%%%%%%%
\begin{minipage}{0.36\textwidth}%
\begin{align*}
\leftrightarrow\frac{1}{2}(\bg^\n P_L)_{ij}\th^{\a\b}\rho_\psi
(T^a)_{st}[-(\bq_\n \bp_\b\\
%%%
-\bq_\b \bp_\n){\bdelta^\m}_\a)-\bk_\a \bp_\b{\bdelta^\m}_\n]
\end{align*}
\end{minipage}
\vskip0.4cm
\begin{minipage}{0.27\textwidth}
\includegraphics[scale=0.6]{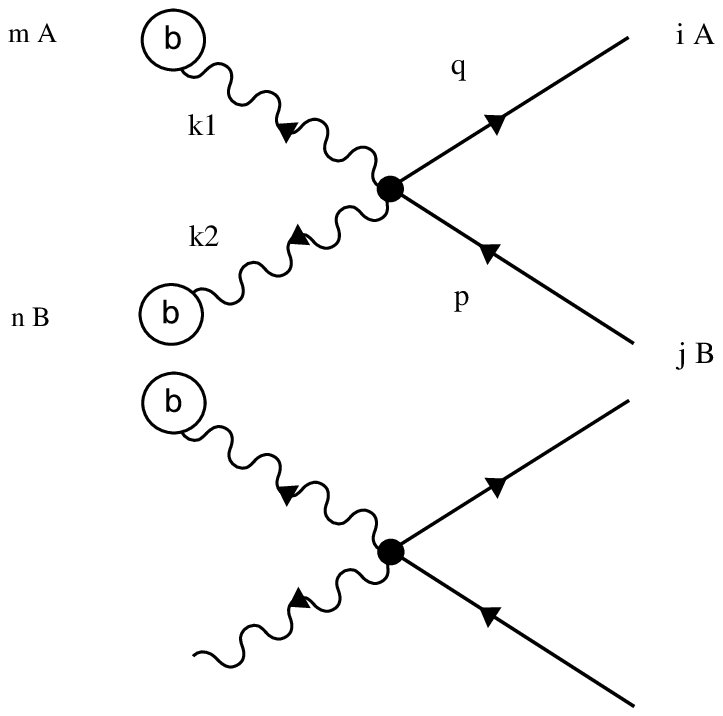}%
\end{minipage}%
\begin{minipage}{0.71\textwidth}%
\begin{align*}
\leftrightarrow&-\frac{1}{4}(\bg^\s P_L)_{ij}\th^{\a\b}\{\rho_\psi(T^a),\rho_\psi(T^b)\}_{st}(-(\bk_1-\bk_2)_\s\bdelta^{\m}_\a\bdelta^\n_\b\\
%%%%
&+2\bk_{1\a}\bdelta^\m_\s\bdelta^\n_\b+2\bk_{2\a}\bdelta^\n_\s\bdelta^\m_\b-(\bq-\bp)_\b(\bdelta^\m_\a\bdelta^\n_\s+\bdelta^\n_\a\bdelta^\m_\s))\\
%%%%
&+[\rho_L(T^i)^{(a)},\rho_L(T^j)^{(b)}]_{st}((\bq+\bp)_\s\bdelta^\m_\a\bdelta^\n_\b-(\bq+\bp)_\b(\bdelta^\m_\a\bdelta^\n_\s-\bdelta^\n_\a\bdelta^\m_\s))].
\end{align*}
\end{minipage}
\caption{Feynman rules of the noncommutative interactions relevant to our calculations, involving the Dirac fermion $\psi$. }
\label{f:1}
\end{figure}

Let us notice that  $\tilde\Gamma^{(n,k)}$ is equivalent to a background 1PI diagram with $n$ background gauge field legs,  $k$ fermionic legs and $k$ anti-fermionic legs. (Note that our definitions do not involve any symmetrisation over the background gauge fields). The vertices relevant to our calculations and their associated Feynman rules for $\alpha=1$  are given in Fig.~\ref{f:1}. In the Feynman rules,  the background field legs are denoted with an encircled ``b"; the rules are defined without symmetrising over background field legs, in accordance with eq.~\eqref{Gammaexp}.

%%%%%%%%%%%%%%%%%%%%%%%%%%%%%%%%%%%%%%%%%%%%%%%%%%%%%%
%%%%%%%%%%%%%%%%%%%%%%%%%%%%%%%%%%%%%%%%%%%%%%%%%%%%%%%	
%%%%%%%%%%%%%%%%%%%%%%%%%%%%%%%%%%%%%%%%%%%%%%%%%%%%%%%	
%%%%%%%%%%%%%%%%%%%%%%%%%%%%%%%%%%%%%%%%%%%%%%%%%%%%%%%

	\section{Computation of the UV divergent part of the effective action involving two fermion fields}
	
	In this section we shall compute the UV divergent contributions (not involving evanescent operators) to the effective action involving two fermion fields, at one-loop and first order in $\theta$, by calculating the background field 1PI diagrams $\tilde\Gamma^{(n,k)}$ with no external quantum field legs of eq.~\eqref{Gammaexp}, using the Feynman rules in Fig.~\ref{f:1}.   
	
 To ease the computation we consider the following simplifications,  which do not mean a loss of generality of the results:
	
\begin{itemize}

\item Choice of gauge $\alpha=1$. This choice greatly simplifies the gauge propagator and since the on-shell effective action is 
independent of the gauge-fixing term --see \cite{Fradkin:1983nw} and references therein--, the conclusions we shall draw 
from our explicit computations upon taking them on-shell will also be gauge independent.

\item Computing a minimum number of diagrams. Since the use of the background field method ensures gauge invariance (modulo one-loop irrelevant evanescent operators) of the result for an anomaly free theory, the full gauge invariant contribution to the UV divergent part 
with no evanescent operator of the effective action --which is local in the fields-- can be reconstructed from a reduced number of 1PI diagrams $\tilde\Gamma^{(n,k)}$. These UV  divergent contributions to the effective action can be expanded in a basis of independent gauge invariant terms. If their contributions with a given number and types of fields are also independent, then the coefficients in the expansion can be fixed by computing the 1PI diagrams with the same number and types of fields.
\end{itemize}

\begin{figure}[h]\centering
\psfrag{a1}{$A_1$}\psfrag{a2}{$A_2$}\psfrag{a3}{$A_3$}
\includegraphics[scale=1.2]{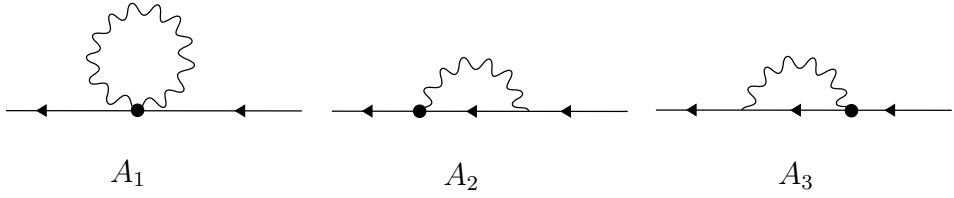}
\caption{Diagrams contributing to  $\tilde\Gamma^{(0,1)}$ at order $h$.}
\label{f:2}
\end{figure}

\begin{figure}[h]\centering
\psfrag{b1}{$B_1$}\psfrag{b2}{$B_2$}\psfrag{b3}{$B_3$}\psfrag{b4}{$B_4$}\psfrag{b5}{$B_5$}\psfrag{b6}{$B_6$}
\psfrag{b7}{$B_7$}\psfrag{b8}{$B_8$}\psfrag{b9}{$B_9$}
\includegraphics[scale=1.2]{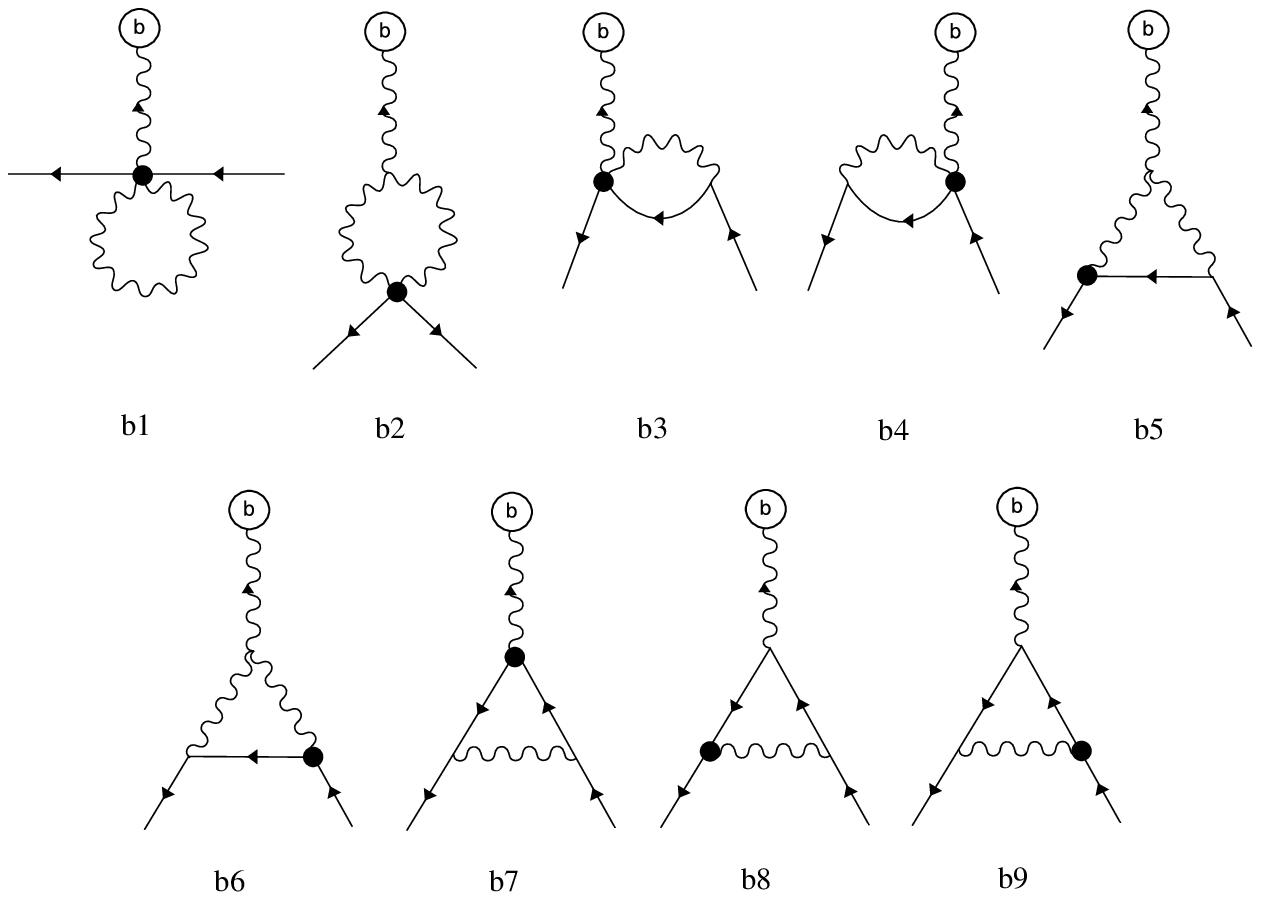}
\caption{Diagrams contributing to  $\tilde\Gamma^{(1,1)}$ at order $h$.}
\label{f:3}
\end{figure}

In order to identify the diagrams that must be computed, we should start by choosing a basis in 4 dimensions of gauge invariant terms whose integrals are independent. Since it was shown in ref.~\cite{Martin:2009sg} that noncommutative GUT inspired theories such as the ones under consideration have no four fermion divergences, a little power-counting takes us to the conclusion that we only need to consider terms with two fermion fields. We choose the following ones, for each flavour $r$:
\begin{align}
\nonumber s^r_1&=\th^{\a\b}\bar\psi_r\ga^\m P_L f_{\m\b}D_\a\psi_r,& s^r_2&=\th^{\a\b}\bar\psi_r\ga^\m P_L f_{\a\b}D_\m\psi_r, & s^r_3&=\th^{\a\b}\bar\psi_r\ga^\m P_L {\cal D}_\m f_{\a\b}\psi_r,\\
%%%
\nonumber s^r_4&=\th^{\a\b}\bar\psi_r\ga_\a P_L f_{\b\m}D^\m\psi_r,& s^r_5&=\th^{\a\b}\bar\psi_r\ga_\a P_L {\cal D}^\m f_{\b\m}\psi_r, &
s^r_6&=\th^{\a\b}\bar\psi_r{\ga_{\a\b}}^\m P_L {\cal D}^\n f_{\m\n}\psi_r,\\
%%%
\nonumber s^r_7&=\th^{\a\b}\bar\psi_r{\ga_{\a\b}}^\m P_L  f_{\m\n}D^\n\psi_r, & s^r_8&=\th^{\a\b}\bar\psi_r{\ga_{\a}}^{\r\s} P_L {\cal D}_\b f_{\r\s}\psi_r, & s^r_9&=\th^{\a\b}\bar\psi_r{\ga_{\a}}^{\r\s} P_L  f_{\r\s}D_\b\psi_r,\\
%%%%
\label{ss}s^r_{10}&=\th^{\a\b}\bar\psi_r{\ga_{\a}}^{\r\s} P_L  f_{\b\s}D_\r\psi_r, & s^r_{11}&=\th^{\a\b}\bar\psi_r\ga_\a D_\b D^2\psi_r, & s^r_{12}&=\th^{\a\b}\bar\psi_r{\ga_{\a\b}}^\m D_\m D^2\psi_r.
\end{align}
In the formulae above, $f_{\m\n}$ and $D_\a f_{\m\n}$ are shorthands for  $\rho_r(f_{\m\n})$ and $\rho_r(D_\a f_{\m\n})$. We will omit explicit indications of the representations $\rho_r$ in future formulae; it will be assumed that a Lie-algebra valued field or generator acting on a fermion $\psi_r$ does so in the representation $\rho_r$.
Note that there are other admissible gauge invariant terms, involving symmetric invariant tensors $t^{a_1\dots a_k}$ of the gauge group, such as $\th^{\a\b} \bar\psi_r \gamma^\m t^{a_1\dots a_k}T_r^{a_1}\dots T_r^{a_{k-1}}(f_{\a\b})^{a_k} D_\m\psi_r$; however, these terms can be seen not to appear in the UV divergent part of the effective action, and may be ignored (also, recall that we are dealing with anomaly safe  theories with $t^{abc}=d^{abc}=0$). Schematically, the terms $s_i$  that appear in the UV divergent part of the effective action are of the form $\theta\bar\psi D^3\psi$ --spanned by $s_{11},s_{12}$, which involve at least two fermion fields and have independent two-field contributions-- and $\theta\bar\psi (Df)\psi, \theta\bar\psi f D\psi$, spanned by $s_1-s_{10}$, which involve at least two fermion fields and a gauge field, in such a way that these contributions are independent of each other. From the discussions above, it is clear that the coefficients of the expansion of the UV divergent contributions to the  effective action involving two fermion fields can be obtained by computing only  1PI diagrams with two fermion fields, $\tilde \Gamma^{(0,1)}$, and with one gauge field and two fermion fields, $\tilde\Gamma^{(1,1)}$. The diagrams are shown in Figs.~\ref{f:2} and \ref{f:3}. Note that there is a subtlety in the computation: the diagrams of Fig.~\ref{f:2} yield the contribution $-i\tilde \Gamma^{(0,1)}_{ij}\hat{\bar\psi}_i\hat\psi_j$  to the effective action (see eq.~\eqref{Gammaexp}), which fixes the coefficients of $s_{11}$ and $s_{12}$ in the expansion of the effective action in terms of the basis of gauge invariant terms. In turn, the diagrams of Fig.~\ref{f:3}  contribute as  $-i\tilde \Gamma^{(1,1)}_{ij(\m,a)}\hat{\bar\psi}_i\hat\psi_j\hat b_\m^a$ to the effective action, 
and this will be a sum of three-field terms coming from both the $s_{11},s_{12}$ combination fixed beforehand and from the $s_1-s_{10}$ terms. In order to get the coefficients of the latter, the three-field contributions  of  $s_{11}$ and $s_{12}$ have to be subtracted.

The results for the UV divergent contributions of the diagrams of figs.~\ref{f:2} and \ref{f:3} are shown in appendix A. The final result for the gauge invariant $O(\th)$ contributions to the divergent part of the one-loop effective action is the following,
\begin{align}\nonumber
&\Gamma^{\rm div}_{\bar\psi\psi}[b_\m,\psi]=\iDx\Big\{\frac{g^2}{192\pi^2\epsilon}\th^{\a\b}\sum_r  C_2(r)\,\bar\psi_r\ga_{\a\b\r}P_L D^\r D^2\psi_r
+\frac{ig^2}{16\pi^2\epsilon}\sum_r C_2(G)\Big[\\
%%%
\nonumber &\frac{1}{6}\th^{\a\b}\bar\psi_r\ga^\m P_Lf_{\m\b}D_\a\psi_r
-\frac{1}{3}\th^{\a\b}\bar\psi_r\ga^\m P_L f_{\a\b}D_\m \psi_r-\frac{1}{8}\th^{\a\b}\bar\psi_r\ga^\m P_L {\cal D}_\m f_{\a\b} \psi_r+\frac{5}{6}\th^{\a\b}\bar\psi_r\ga_\a P_L  f_{\b\m} D^\m\psi_r\\
%%%
\nonumber &+\frac{5}{12}\th^{\a\b}\bar\psi_r\ga_\a P_L{\cal D}^\m f_{\b\m}  \psi_r-\frac{1}{8}\th^{\a\b}\bar\psi_r{\ga_\a}^{\r\s} P_L {\cal D}_\b f_{\r\s} \psi_r-\frac{1}{16}\th^{\a\b}\bar\psi_r{\ga_{\a\b}}^{\m} P_L {\cal D}^\n f_{\m\n} \psi_r\Big]\\
%%%%
%%%%%%
\nonumber &+\frac{ig^2}{16\pi^2\epsilon}\sum_r C_2(r)\Big[\frac{1}{2}\th^{\a\b}\bar\psi_r\ga^\m P_Lf_{\m\b}D_\a\psi_r+\frac{1}{8}\th^{\a\b}\bar\psi_r\ga^\m P_L {\cal D}_\m f_{\a\b} \psi_r-\frac{3}{2}\th^{\a\b}\bar\psi_r\ga_\a P_L  f_{\b\m} D^\m\psi_r\\
%%%
\nonumber &-\frac{3}{4}\th^{\a\b}\bar\psi_r\ga_\a P_L{\cal D}^\m f_{\b\m}  \psi_r+\frac{1}{24}\th^{\a\b}\bar\psi_r{\ga_\a}^{\r\s} P_L {\cal D}_\b f_{\r\s} \psi_r+\frac{1}{6}\th^{\a\b}\bar\psi_r{\ga_{\a\b}}^{\m} P_L {\cal D}^\n f_{\m\n} \psi_r\\
%%%
\label{Gammadiv}&+\frac{1}{12}\th^{\a\b}\bar\psi_r{\ga_{\a\b}}^{\m} P_L  f_{\m\n} D^\n \psi_r\Big]\Big\},
\end{align}
where all covariant derivatives and field strengths are evaluated on the background field $b_\m$, and we have suppressed the hats on the classical fields as well as explicit indications of the representations $\rho_r$ of the field strength and its covariant derivatives to ease the notation.
In the formulae above, $C_2(r)$ represents the second Casimir of the representation $r$, $C_2(G)$ corresponding to the adjoint representation. It is defined as $T^a_r T^a_r=C_2(r)\mathbb{I}_r$. In terms of the basis $s_i$ of eq.~\eqref{ss}
\begin{align}
\nonumber&\Gamma^{\rm div}_{\bar\psi\psi}[b_\m,\psi]=\idx\Big(\frac{g^2i}{16\pi^2\epsilon}\sum_r C_2(G)\Big[\frac{1}{6}s^r_1-\frac{1}{3}s^r_2-\frac{1}{8}s^r_3+\frac{5}{6}s^r_4+\frac{5}{12}s^r_5-\frac{1}{16}s^r_6-\frac{1}{8}s^r_8\Big]\\
%%%
\label{Gammadivs}
&+\frac{g^2i}{16\pi^2\epsilon}\sum_r C_2(r)\Big[\frac{1}{2}s^r_1+\frac{1}{8}s^r_3-\frac{3}{2}s^r_4-\frac{3}{4}s^r_5+\frac{1}{6}s^r_6+\frac{1}{12}s^r_7+\frac{1}{24}s^r_8-\frac{i}{12}s^r_{12}\Big]\Big).
\end{align}
%%%%%%%%%%%%%%%%%%%%%%%%%%%%%%%%%%%%%%%%%%%%%%%%%%%%%%%
%%%%%%%%%%%%%%%%%%%%%%%%%%%%%%%%%%%%%%%%%%%%%%%%%%%%%%%
%%%%%%%%%%%%%%%%%%%%%%%%%%%%%%%%%%%%%%%%%%%%%%%%%%%%%%%	

	\section{Analysing renormalisability of the matter sector}
	
	The objective of this section is to check whether the noncommutative divergences involving two fermion fields of eq.~\eqref{Gammadiv} can be subtracted by means of multiplicative renormalisation of the coupling constant, the noncommutative parameter $\theta$ and fields plus counterterms which vanish on-shell.
	
	First, the  the one-loop divergences at order zero in $\theta$ are, as is well known, renormalisable by means of multiplicative renormalisations of fields and parameters. These multiplicative renormalisations take the form
\begin{align}\label{multipren}
 b_\m&=  Z_b^{1/2}b_\m^{R}, & \psi&= Z_\psi^{1/2}\psi^{R},& g&=\mu^{-\epsilon}\,Z_g g^R, & \theta_{\m\n}&=Z_\th \theta^R_{\m\n},
\end{align}
with $Z_i=1+\delta Z_i$. The parameter $\mu$ is the dimensional regularisation scale. Note that in the background field method there is no need to renormalise the quantum field $q$ of eq.~\eqref{splitting}. It is easily seen that gauge invariance forces $\delta Z_b=0$, while the divergences in the ordinary theory yield
\begin{align}\label{Zs}
\delta Z^r_\psi&=\frac{g^2 C_2(r)}{16\pi^2\epsilon},\\
%%%
\nonumber
\delta Z_g&=\frac{g^2}{16\pi^2\epsilon}\Big[\frac{11}{6}C_2(G)-\frac{2}{3}\sum_r c_2(r)\Big],
\end{align}
where $c_2(r)$ is the index of the representation $r$; its relation with $C_2(r)$ is given in eq.~\eqref{Cs}.
 Let us introduce the following type of counterterms
\begin{equation}\label{redundant}
S^{\rm ct}=\iDx \frac{\delta S}{\delta a_\m^a(x)} F_\m^a[a,\psi]+\Big(\sum_r\frac{\delta S}{\delta\psi_{r}(x)}G_r[a,\psi]+{\rm c.c.}\Big),
\end{equation}
which vanish on-shell due to the equations of motion
\begin{align*}
\frac{\delta S}{\delta a_\m^a(x)}=\frac{\delta S}{\delta\psi_r(x)}=0.
\end{align*}
In order to preserve gauge symmetry,  $F^a[a,\psi]$ and $G_r[a,\psi]$ have to transform in 4 dimensions under gauge transformations as follows 
 \begin{align*}
s F_\m=-i[F_\m,\la],\quad sG_r=i\lambda G_r.
\end{align*}
 We consider the following $F_\m$ and $G_{r,L}$
\begin{align}
\nonumber F_\m=&y_1\th^{\a\b}{\cal D}_\m f_{\a\b}+y_2{\th_\m}^\a{\cal D}^\n f_{\n\a}+\sum_r y^r_3{\th_\m}^\a(\bar\psi_r\ga_\a P_L T^a\psi_r)T^a\\
%%%
\nonumber &+i\sum_r y^r_4{\th}^{\a\b}(\bar\psi_r\ga_{\m\a\b} P_L T^a\psi_r)T^a+y_5{\tilde{\th}_\m}^{\,\,\,\,\b} D^\n f_{\n\b},\\
%%%
\nonumber G_{r,L}=&k^r_1\th^{\a\b}f_{\a\b}P_L\psi_r+k^2_r\th^{\a\b}{\ga_{\a\m}}P_L{f_\b}^\m \psi_r+k^r_3\th^{\a\b}\ga_{\a\m}P_L D_\b D^\m\psi_r+k^r_4\th^{\a\b}\ga_{\a\b}P_L D^2\psi_r\\
%%%
&+k^r_5\tilde{\th}^{\a\b}\ga_5 P_L f_{\a\b}\psi_r; \,\,\,y_i\in\mathbb{R},\,k_i\in\mathbb{C},
\label{redundant2}
\end{align}
which have the appropriate behaviour under gauge transformations.
Note that we only considered a left-handed part for $G_r$, since all the divergences in eq.~\eqref{Gammadiv} involve left-handed projectors $P_L$.

The $O(\th)$ counterterm action involving two fermion fields obtained by considering the multiplicative renormalisations of eq.~\eqref{multipren} and the tree-level contributions of the counterterms vanishing on-shell of eqs.~\eqref{redundant} and \eqref{redundant2}, has the following expansion in the basis of terms $s_i$:
\begin{align}
S^{\rm ct}_{\bar\psi\psi}=ih\iDx\sum_{r,i} C_i^r s^r_i,\label{ct}
\end{align}
\begin{align*}
C^r_1&=\frac{1}{2}(\delta Z_\th+\delta Z_\psi)-(k^r_2)^*-k^r_2, & C^r_2&=-\frac{1}{4}(\delta Z_\th+\delta Z_\psi)+(k^r_1)^*+k^r_1+\frac{i}{2}(k^r_3)^*-\frac{i}{2}k^r_3,\\
%%%%
C^r_3&=-i y_1+k^r_1-\frac{1}{2}k^r_2+\frac{i}{2}(k^r_3)^*, & C^r_4&=-(k^r_2)^*-k^r_2-i(k^r_3)^*-ik^r_3-4i(k^r_4)^*,\\
%%%%
C^r_5&=iy_2-\frac{i}{2g^2} y^r_3-2i(k^r_4)^*-k^r_2, & C^r_6&=-\frac{1}{2g^2}y^r_4-\frac{1}{2}y_5-i(k^r_4)^*-ik^r_5,\\
%%%
C^r_7&=-2i(k^r_4)^*-i(k^r_5)^*-ik^r_5, & C^r_8&=-\frac{1}{2}k^r_2+\frac{i}{2}(k^r_3)^*+\frac{i}{2}k^r_3-ik^r_5,\\
%%%%
C^r_9&=\frac{i}{2}(k^r_3)^*+\frac{i}{2}k^r_3-ik^r_5-i(k^r_5)^*, & C^r_{10}&=(k^r_2)^*-k^r_2+i(k^r_3)^*+ik^r_3,\\
%%%%
C^r_{11}&=-(k^r_3)^*-k^r_3-2(k^r_4)^*-2k^r_4,& C^r_{12}&=-(k^r_4)^*+k^r_4.
\end{align*}
The  $y_i,k_i$ of eq.~\eqref{redundant2}  also generate, to $O(\th)$  and at tree-level, terms involving four fermion fields,
\begin{align}\label{ct2}
S^{\rm ct}_{\bar\psi\psi\bar\psi\psi}=\idx\sum_{r,s}(y^s_3\th^{\a\b}+2y^s_4\tilde\th^{\a\b})(\bar\psi_r\ga_\a T^a P_L\psi_r)(\bar\psi_s\ga_\b T^bP_L\psi_s),\,\,\tilde\th^{\a\b}=\frac{1}{2}\epsilon^{\a\b\m\n}\th_{\m\n}.
\end{align}
As seen in ref.~\cite{Martin:2009sg}, four-fermion divergences are absent in noncommutative GUT compatible theories at one loop, $O(\theta)$. Renormalisability of the divergences involving two and four fermions amounts to demand
\begin{align*}
S^{\rm ct}_{\bar\psi\psi\bar\psi\psi}=0,\quad S^{\rm ct}_{\bar\psi\psi}=-\Gamma^{\rm div}_{\bar\psi\psi}.
\end{align*}
where $S^{\rm ct}_{\bar\psi\psi\bar\psi\psi},\,S^{\rm ct}_{\bar\psi\psi}$ and $\Gamma^{\rm div}_{\bar\psi\psi}$ are given, respectively, by eqs.~\eqref{ct2}, \eqref{ct} and \eqref{Gammadivs}
The first equation is solved by choosing $y^r_3,y^r_4$ to be flavour independent ($y_i^r=y_i\,\forall\, r$). Solving the second identity in the basis of independent terms $s^r_i$ and projecting the resulting equations into their real and imaginary parts, one gets
\begin{align*}
&s^r_1:  & &\frac{1}{2}(\delta Z_\th+\delta Z_\psi)-2{\rm Re}k^r_2=-\frac{g^2}{16\pi^2\epsilon}\Big(\frac{1}{6}C_2(G)+\frac{1}{2}C_2(r)\Big),\\
%%%
&s^r_2:  &&-\frac{1}{4}(\delta Z_\th+\delta Z_\psi)+2{\rm Re}k^r_1+{\rm Im}k^r_3=\frac{g^2C_2(G)}{48\pi^2\epsilon},\\
%%%
&s^r_3: &&y_1-\frac{1}{2}{\rm Re}k^r_3-{\rm Im}k^r_1+\frac{1}{2}{\rm Im}k^r_2=0, \quad \frac{1}{2}{\rm Im}k^r_3+{\rm Re}k^r_1-\frac{1}{2}{\rm Re}k^r_2=\frac{g^2(C_2(G)-C_2(r))}{128\pi^2\epsilon},\\
%%%
&s^r_4: && {\rm Re}k^r_3+2{\rm Re}k^r_4=0,\quad-2{\rm Re}k^r_2-4{\rm Im}k^r_4=-\frac{g^2}{16\pi^2\epsilon}\Big(\frac{5}{6}C_2(G)-\frac{3}{2}C_2(r)\Big), \\
%%%%
&s^r_5: && \frac{1}{2g^2}y^r_3-y_2+2{\rm Re}k^r_4+{\rm Im}k^r_2=0,\quad-{\rm Re}k^r_2-2{\rm Im}k^r_4=-\frac{g^2}{16\pi^2\epsilon}\Big(\frac{5}{12}C_2(G)-\frac{3}{4}C_2(r)\Big), 
\end{align*}
\begin{align*}
%%%%
&s^r_6: && {\rm Re}k^r_4+{\rm Re}k^r_5=0,\quad-\frac{1}{2g^2}y^r_4-\frac{1}{2}y_5-{\rm Im}k^r_4+{\rm Im}k^r_5=-\frac{g^2}{16\pi^2\epsilon}\Big(-\frac{1}{16}C_2(G)+\frac{1}{6}C_2(r)\Big), \\
%%%
&s^r_7: && {\rm Re}k^r_4+{\rm Re}k^r_5=0,\quad-2{\rm Im}k^r_4=-\frac{g^2C_2(r)}{192\pi^2\epsilon}, \\
%%%%
&s^r_8:&& -{\rm Re}k^r_3+{\rm Re}k^r_5+\frac{1}{2}{\rm Im}k^r_2=0,\quad{\rm Im}k^r_5-\frac{1}{2}{\rm Re}k^r_2=-\frac{g^2}{16\pi^2\epsilon}\Big(-\frac{1}{8}C_2(G)+\frac{1}{24}C_2(r)\Big), \\
%%%
&s_9^r: && -{\rm Re}k^r_3+2{\rm Re}k^r_5=0,\\
%%%
&s^r_{10}: && {\rm Im}k^r_2-{\rm Re}k^r_3=0,\\
%%%
&s^r_{11}: && -{\rm Re}k^r_3-2{\rm Re}k^r_4=0,\\
%%%
 &s^r_{12}: && -2{\rm Im}k^r_4=-\frac{g^2C_2(r)}{192\pi^s\epsilon}.
\end{align*}
The equations  are compatible, and we find the following family of solutions
\begin{align}
\nonumber y_1&={\rm Im}k^r_1,& y^r_3&=2g^2 y_2,\\
%%%%
\nonumber y^r_4&=-y_5 g^2-\frac{g^4}{384\pi^2}(16 C_2(r)-13 C_2(G)),&
 Z_\th&=-Z_\psi-\frac{g^2}{48\pi^2\epsilon}(13C_2(r)-4C_2(G)), \\
 %%%
 \nonumber {\rm Re}k^r_1&=-\frac{1}{2}{\rm Im}k^r_3-\frac{g^2}{384\pi^2\epsilon}(13C_2(r)-8C_2(G)), & {\rm Im}k^r_5&=-\frac{g^2}{384\pi^2\epsilon}(11C_2(r)-8C_2(G)),\\
  %%%
 \nonumber  {\rm Im}k^r_4&=\frac{g^2 C_2(r)}{384\pi^2\epsilon},& {\rm Re}k^r_2&=-\frac{5g^2}{192\pi^2\epsilon}(2C_2(r)-C_2(G)),\\
  %%%
{\rm Im}k^r_2&={\rm Re}k^r_3=2{\rm Re}k^r_5=-2{\rm Re}k^r_4.\label{sols}
  %%%
\end{align}
First, note that  $y_1,y_2,y_5$ and $\delta Z_\th$ must be flavour independent  (see eq.~\eqref{redundant2}), and so must be $y_3,y_4$ for the cancellation of the four fermion divergences to be preserved under the renormalisation procedure, as was previously seen. Looking at the solutions in eq.~\eqref{sols} and imposing flavour independence, it is clear that one must require that all flavours have identical  $C_2(r)$; this can be achieved by considering all fields in the same representation or also in its conjugate.

Since the counterterms  eqs.~\eqref{redundant} and \eqref{redundant2}, which are dependent on the parameters $y_i,$ and $k_i$, vanish on-shell, we have that the corresponding divergences in the effective action that they are able to subtract are physically
irrelevant since they will cancel out  when computing the S matrix. Looking at the expansion in the counterterm action of eq.~\eqref{ct}, it is clear that the only divergences surviving on-shell are those associated with the multiplicative renormalisation of the gauge coupling constant and noncommutativity parameter $\theta$.
%%%%%%%%%%%%%%%%%%%%%%%%%%%%%%%%%%%%%%%%%%%%%%%%%%%%%%%
%%%%%%%%%%%%%%%%%%%%%%%%%%%%%%%%%%%%%%%%%%%%%%%%%%%%%%%
%%%%%%%%%%%%%%%%%%%%%%%%%%%%%%%%%%%%%%%%%%%%%%%%%%%%%%%
	
	\section{No $O(\theta)$ UV divergent contributions in the gauge sector}
	
	The results of the previous section, together with those of ref.~\cite{Martin:2009sg}, show that the matter sector of the one-loop, order $\theta$ effective action is renormalisable. It remains to see if the gauge sector is also renormalisable. In all cases analysed in the literature so far, the gauge sector of noncommutative gauge theories in the enveloping algebra approach turned out to be one-loop renormalisable at order $\theta$. We show in this section that, at one-loop, 
there are no order one in $\theta$ UV divergent contributions to the part of the  background-field effective action,  
$\Gamma[\hat b_\m,\hat \psi=0,\hat{\bar\psi}=0]$ which only depends on the gauge field. 
	
	Possible UV divergences in the gauge sector can be of two types, depending on whether they involve $\epsilon$ tensors or not. Since any vector-like contribution to the effective action can always be regularised in a gauge-invariant way in the framework of dimensional regularisation, the allowed $O(\th)$ vector-like UV divergences can only be a combination of the terms $\Tr\theta^{\a\b}f_{\a\b}f_{\m\n}f^{\m\n}$ and $\Tr\theta^{\a\b}f_{\a\m}f_{\b\n}f^{\m\n}$, which vanish for anomaly safe groups since they involve vanishing symmetrised $\Tr T^a\{T^b,T^c\}$. It only remains to show that there are no UV divergences involving $\epsilon$ tensors. These divergences would come from fermionic loops, since the $\epsilon$ tensors arise from traces of $\gamma$ matrices. The one-loop fermionic contributions to the gauge sector of the effective action can be computed in a clever way, to all orders in $\theta$, using the technique used in refs.~\cite{Brandt:2003fx}~and~\cite{Martin:2007wv}. By defining appropriately the dimensionally regularised interactions  --recall that there is an infinity of choices, differing by evanescent contributions-- a change of variables can be done in the fermionic path integral which amounts to inverting the SW map and whose Jacobian is unity. The diagrams to compute involve vertices with noncommutative fermions, in which the noncommutative phase factors are independent of the loop momenta. Then, as done in ref.~\cite{Brandt:2003fx}, it can be easily seen that these diagrams have vanishing UV divergent contributions involving $\epsilon$ tensors.
	
	We have seen  that there are no  order one in $\theta$ UV divergences contributions in the gauge sector at one-loop. 
Now, since at tree-level there are no $O(\th)$ contributions only involving gauge fields, there is no conflict with the multiplicative renormalisations of eq.~\eqref{Zs}, and thus the gauge sector is one-loop renormalisable up to first order in
$\theta$.

%%%%%%%%%%%%%%%%%%%%%%%%%%%%%%%%%%%%%%%%%%%%%%%%%%%%%%%
%%%%%%%%%%%%%%%%%%%%%%%%%%%%%%%%%%%%%%%%%%%%%%%%%%%%%%%
%%%%%%%%%%%%%%%%%%%%%%%%%%%%%%%%%%%%%%%%%%%%%%%%%%%%%%%	
	\section{Summary, conclusions and outlook}
	
	In this paper we have computed the UV divergent contributions, involving two fermions and an arbitrary number of 
gauge fields, to the background field effective action of noncommutative, anomaly safe GUT inspired theories with no scalars. 
We have done the computation at one-loop and first order in the noncommutativity parameter $\theta$. We have shown that those UV 
divergences can be renormalised by means of the ordinary multiplicative renormalisations of the coupling constant and fields,  
along with a multiplicative renormalisation of the noncommutative parameter $\theta$ and with the introduction of $\theta$-dependent 
counterterms which vanish on-shell, provided the irreps furnished by the matter fermionic fields share the same second-degree 
Casimir  invariant. It is obvious that this condition on the second-degree Casimir invariant is automatically fulfilled by the 
fermionic matter content of the phenomenologically relevant ordinary SO(10) and ${\rm E}_6$ GUTs.
We have also shown that the gauge sector of these theories receives no linear one-loop UV divergent radiative corrections which are
of order one in $\theta$.

    Our results, together with those of ref.~\cite{Martin:2009sg} proving the absence of 4 fermion UV divergences in the one-loop effective action of noncommutative GUT inspired theories at first order in $\theta$, show that the theories considered in this paper are renormalisable  
on-shell at one-loop and first order in $\theta$. We have thus seen  that, at one-loop and first order in $\theta$, only the 
renormalisation of the coupling constant and the noncommutative parameter $\theta$ --the two free physical parameters of the classical theory-- are needed to obtain renormalised S matrix elements. This is the first time in which a noncommutative gauge theory defined by means of Seiberg-Witten map, with fermions in representations other than the adjoint, has been shown to have this property.  This result clearly favours the consideration of GUT compatible noncommutative theories over their nonrenormalisable brethren. The only other known examples of  one-loop, $O(\th)$ renormalisable noncommutative gauge theories involve $SU(N)$ adjoint Majorana fermions in a supersymmetric setting~\cite{Martin:2009mu}.
    
  A pressing open problem is the study, at  one-loop and first order 
in $\theta$, of the renormalisability of the noncommutative GUT theories obtained by adding a  noncommutative Higgs and Yukawa sectors 
--through the hybrid Seiberg-Witten map of ref. \cite{Aschieri:2002mc}-- to the noncommutative theories  considered here. 
The computations involved in this study are far more lengthy that the already long calculations carried out in this paper 
and  will certainly deserve to be the content of a different paper.  We hope that the results presented here will encourage 
people to further analyse the properties of noncommutative GUT theories.
	
%%%%%%%%%%%%%%%%%%%%%%%%%%%%%%%%%%%%%%%%%%%%%%%%%%%%%%%
%%%%%%%%%%%%%%%%%%%%%%%%%%%%%%%%%%%%%%%%%%%%%%%%%%%%%%%
%%%%%%%%%%%%%%%%%%%%%%%%%%%%%%%%%%%%%%%%%%%%%%%%%%%%%%%
\section{Acknowledgements}	 
%%%%%%%%%%%%%%%%%%%%%%%%%%%%%%%%%%%%%%%%%%%%%%%%%%%%%%%
%%%%%%%%%%%%%%%%%%%%%%%%%%%%%%%%%%%%%%%%%%%%%%%%%%%%%%%
%%%%%%%%%%%%%%%%%%%%%%%%%%%%%%%%%%%%%%%%%%%%%%%%%%%%%%%	
	This work has been financially supported in part by MICINN through grant
FPA2008-04906, and by the National Science Foundation under Grant No. PHY05-51164. The work of C.~Tamarit has also received financial support from MICINN and the Fulbright Program through grant 2008-0800.
	
\appendix
\section{Divergent contributions to the diagrams of Figs.~\ref{f:2} and \ref{f:3}}
Here we give the pole part of the Feynman diagrams depicted in figs. \ref{f:2} and \ref{f:3}, computed in dimensional regularisation with $D=4+2\epsilon$ dimensions.
\begin{align*}
A_1=&0,\\
%%%
A_2=&-\frac{g^2}{96\pi^2\epsilon}\bigoplus_r C_2(r)\mathbb{I}_r\th^{\a\b}\ga_\b P_L p^2 p_\a -\frac{g^2}{384\pi^2\epsilon}\bigoplus_r C_2(r)\mathbb{I}_r\th^{\a\b}\ga_{\r\a\b} P_L p^2 p^\r,\\
%%%
A_3=&\frac{g^2}{96\pi^2\epsilon}\bigoplus_r C_2(r)\mathbb{I}_r\th^{\a\b}\ga_\b P_L p^2 p_\a -\frac{g^2}{384\pi^2\epsilon}\bigoplus_r C_2(r)\mathbb{I}_r\th^{\a\b}\ga_{\r\a\b} P_L p^2 p^\r,\\
%%%
B_{1}=&0,\\
B_{2}=&-\frac{g^2}{8\pi^2\epsilon}\bigoplus_r C_2(G)T^A_r(\ga^\m k_\a p_\b\th^{\a\b}-\pslash k_\a\th^{\a\m}+\kslash p_\a\th^{\m\n})P_L,\\
%%%
%%%%
%%%%
B_3=&-\frac{g^2}{8\pi^2\epsilon}\bigoplus_r C_2(r)T^A_r\Big(\frac{1}{8}\ga^\m k_\a p_\b\th^{\a\b}+\frac{1}{8}p^\m k_\a\ga_\b\th^{\a\b}+\frac{1}{24}p^\m p_\a \ga_\b\th^{\a\b}+\frac{1}{8}\kslash p_\a\th^{\a\m}+\frac{1}{24}\pslash p_\a\th^{\a\m}\\
&-\frac{1}{8}k\cdot p\ga_\b\th^{\m\b}-\frac{1}{8}p^2\ga_\b\th^{\m\b}+\frac{1}{8}p^\r k_\a{\ga_{\r\b}}^\m\th^{\a\b}+\frac{1}{24}p^\r p_\a {\ga_{\r\b}}^\m\th^{\a\b}+\frac{1}{48}p^2{\ga^\m}_{\a\b}\th^{\a\b}\\
&-\frac{1}{8}k^\r p^\s \ga_{\r\s\b}\th^{\m\b}\Big)P_L\\
%%%
%%%
&-\frac{g^2}{32\pi^2\epsilon}\bigoplus_r C_2(G)T^A_r\Big(-\frac{1}{4}p^\m p_\a \ga_\b\th^{\a\b}-\frac{1}{4}\pslash p_\a\th^{\a\m}+\frac{1}{4}p^2\ga_\b\th^{\m\b}-\frac{1}{4}p^\r p_\a {\ga_{\r\b}}^\m\th^{\a\b}\Big)P_L,\\
%%%%%
%%%%
%%%%
B_{4}=&-\frac{g^2}{8\pi^2\epsilon}\bigoplus_F C_2(F)T^A_F\Big(\frac{1}{8}\ga^\m k_\a p_\b\th^{\a\b}-\frac{1}{6}k^\m k_\a\ga_\b\th^{\a\b}+\frac{1}{6}p^\m k_\a\ga_\b\th^{\a\b}-\frac{1}{6}\kslash k_\a \th^{\a\m}+\frac{1}{24}\pslash k_\a\th^{\a\m}\\
%%%%
&+\frac{1}{24}k^\m p_\a \ga_\b\th^{\a\b}-\frac{1}{24}p^\m p_\a \ga_\b\th^{\a\b}+\frac{1}{6}\kslash p_\a\th^{\a\m}-\frac{1}{24}\pslash p_\a\th^{\a\m}+\frac{1}{4}k^2\ga_\b\th^{\m\b}-\frac{3}{8}k\cdot p\ga_\b\th^{\m\b}\\
%%%
&+\frac{1}{8}p^2\ga_\b\th^{\m\b}+\frac{1}{6}k^\r k_\a{\ga_{\r\b}}^\m\th^{\a\b}-\frac{1}{6}p^\r k_\a{\ga_{\r\b}}^\m\th^{\a\b}-\frac{1}{24}k^\r p_\a{\ga_{\r\b}}^\m\th^{\a\b}+\frac{1}{24}p^\r p_\a {\ga_{\r\b}}^\m\th^{\a\b}\\
&+\frac{1}{48}k^2{\ga^\m}_{\a\b}\th^{\a\b}-\frac{1}{24}k\cdot p{\ga^\m}_{\a\b}\th^{\a\b}+\frac{1}{48}p^2{\ga^\m}_{\a\b}\th^{\a\b}+\frac{1}{8}k^\r p^\s \ga_{\r\s\b}\th^{\m\b}\Big)P_L\\
%%%
%%%
&-\frac{g^2}{32\pi^2\epsilon}\bigoplus_F C_2(G)T^A_F\Big(\frac{1}{4}k^\m k_\a\ga_\b\th^{\a\b}-\frac{1}{4}p^\m k_\a\ga_\b\th^{\a\b}+\frac{1}{4}\kslash k_\a \th^{\a\m}-\frac{1}{4}\pslash k_\a\th^{\a\m}\\
%%%%
&-\frac{1}{4}k^\m p_\a \ga_\b\th^{\a\b}+\frac{1}{4}p^\m p_\a \ga_\b\th^{\a\b}-\frac{1}{4}\kslash p_\a\th^{\a\m}+\frac{1}{4}\pslash p_\a\th^{\a\m}-\frac{1}{4}k^2\ga_\b\th^{\m\b}+\frac{1}{2}k\cdot p\ga_\b\th^{\m\b}\\
%%%
&-\frac{1}{4}p^2\ga_\b\th^{\m\b}-\frac{1}{4}k^\r k_\a{\ga_{\r\b}}^\m\th^{\a\b}+\frac{1}{4}p^\r k_\a{\ga_{\r\b}}^\m\th^{\a\b}+\frac{1}{4}k^\r p_\a{\ga_{\r\b}}^\m\th^{\a\b}-\frac{1}{4}p^\r p_\a {\ga_{\r\b}}^\m\th^{\a\b}\Big)P_L,\\
\end{align*}
%%%%%%%%%%%%%%%%
%%%%%%%%%%%%%%%%
%%%%%%%%%%%%%%%%
\begin{align*}
B_5=&\frac{g^2}{32\pi^2\epsilon}\bigoplus_r C_2(G)T^A_r\Big(\frac{5}{6}\ga^\m k_\a p_\b\th^{\a\b}-\frac{1}{6}k^\m k_\a\ga_\b\th^{\a\b}+\frac{1}{6}p^\m k_\a\ga_\b\th^{\a\b}-\frac{1}{3}\kslash k_\a \th^{\a\m}-\frac{1}{3}\pslash k_\a\th^{\a\m}\\
%%%%
&-\frac{1}{6}k^\m p_\a \ga_\b\th^{\a\b}+\frac{1}{6}p^\m p_\a \ga_\b\th^{\a\b}+\frac{5}{6}\kslash p_\a\th^{\a\m}-\frac{1}{6}\pslash p_\a\th^{\a\m}+\frac{1}{3}k^2\ga_\b\th^{\m\b}-\frac{1}{2}k\cdot p\ga_\b\th^{\m\b}\\
%%%
&+\frac{1}{6}p^2\ga_\b\th^{\m\b}-\frac{1}{3}k^\r k_\a{\ga_{\r\b}}^\m\th^{\a\b}-\frac{1}{6}p^\r k_\a{\ga_{\r\b}}^\m\th^{\a\b}+\frac{2}{3}k^\r p_\a{\ga_{\r\b}}^\m\th^{\a\b}-\frac{1}{6}p^\r p_\a {\ga_{\r\b}}^\m\th^{\a\b}\\
&-\frac{1}{24}k^\m k^\r {\ga_{\r\a\b}}\th^{\a\b}+\frac{1}{12}p^\m k^\r {\ga_{\r\a\b}}\th^{\a\b}+\frac{1}{24}k^\m p^\r {\ga_{\r\a\b}}\th^{\a\b}-\frac{1}{12}p^\m p^\r {\ga_{\r\a\b}}\th^{\a\b}\\
%%%
&+\frac{1}{6}k^\r p^\s \ga_{\r\s\b}\th^{\m\b}\Big)P_L,\\
%%%%%%%%%%%%%%%%
%%%%%%%%%%%%%%%%
%%%%%%%%%%%%%%%%
%%%%%%%%%%%%%%%%
%%%%%%%%%%%%%%%%
%%%%%%%%%%%%%%%%
B_6=&\frac{g^2}{32\pi^2\epsilon}\bigoplus_r C_2(G)T^A_r\Big(\frac{5}{6}\ga^\m k_\a p_\b\th^{\a\b}+\frac{1}{3}p^\m k_\a\ga_\b\th^{\a\b}-\frac{1}{2}\pslash k_\a \th^{\a\m}-\frac{1}{6}p^\m p_\a \ga_\b\th^{\a\b}+\frac{2}{3}\kslash p_\a\th^{\a\m}\\
&+\frac{1}{6}\pslash p_\a\th^{\a\m}-\frac{1}{6}k\cdot p\ga_\b\th^{\m\b}-\frac{1}{6}p^2\ga_\b\th^{\m\b}
+\frac{1}{3}p^\r k_\a{\ga_{\r\b}}^\m\th^{\a\b}-\frac{1}{2}k^\r p_\a {\ga_{\r\b}^\m}\th^{\a\b}-\frac{1}{6}p^\r p_\a {\ga_{\r\b}}^\m\th^{\a\b}\\
%%%
&+\frac{1}{24}k^\m p^\r {\ga_{\r\a\b}}\th^{\a\b}-\frac{1}{12}p^\m p^\r {\ga_{\r\a\b}}\th^{\a\b}-\frac{1}{6}k^\r p^\s \ga_{\r\s\b}\th^{\m\b}\Big)P_L,\\
%%%%%%%%%%%%%%%%
%%%%%%%%%%%%%%%%
%%%%%%%%%%%%%%%%
%%%%%%%%%%%%%%%%
%%%%%%%%%%%%%%%%
%%%%%%%%%%%%%%%%
B_7=&-\frac{g^2}{16\pi^2\epsilon}\bigoplus_r\Big(C_2(r)-\frac{1}{2}C_2(G)\Big)T^A_r\Big(-\frac{1}{3}\ga^\m k_\a p_\b\th^{\a\b}-\frac{1}{12}k^\m k_\a\ga_\b\th^{\a\b}+\frac{1}{6}p^\m k_\a\ga_\b\th^{\a\b}\\
%%%
&+\frac{1}{4}\kslash k_\a \th^{\a\m}-\frac{1}{6}\pslash k_\a\th^{\a\m}-\frac{1}{3}\kslash p_\a\th^{\a\m}+\frac{1}{12}k^2\ga_\b\th^{\m\b}-\frac{1}{6}k\cdot p\ga_\b\th^{\m\b}
-\frac{1}{4}k^\r k_\a{\ga_{\r\b}}^\m\th^{\a\b}\Big)P_L,\\
%%%%%%%%%%%%%%%%
%%%%%%%%%%%%%%%%
%%%%%%%%%%%%%%%%
B_8=&-\frac{g^2}{16\pi^2\epsilon}\bigoplus_F\Big(C_2(F)-\frac{1}{2}C_2(G)\Big)T^A_F\Big(\frac{1}{6}\ga^\m k_\a p_\b\th^{\a\b}-\frac{1}{3}k^\m k_\a\ga_\b\th^{\a\b}+\frac{7}{12}p^\m k_\a\ga_\b\th^{\a\b}\\
%%%
&-\frac{1}{6}\kslash k_\a \th^{\a\m}+\frac{1}{12}\pslash k_\a\th^{\a\m}+\frac{1}{6}k^\m p_\a \ga_\b\th^{\a\b}-\frac{5}{12}p^\m p_\a \ga_\b\th^{\a\b}+\frac{1}{6}\kslash p_\a\th^{\a\m}-\frac{1}{12}\pslash p_\a\th^{\a\m}\\
%%%
&+\frac{1}{6}k^2\ga_\b\th^{\m\b}-\frac{1}{4}k\cdot p\ga_\b\th^{\m\b}+\frac{1}{12}p^2\ga_\b\th^{\m\b}-\frac{1}{6}k^\r k_\a{\ga_{\r\b}}^\m\th^{\a\b}-\frac{1}{12}p^\r k_\a{\ga_{\r\b}}^\m\th^{\a\b}\\
%%%
&+\frac{1}{3}k^\r p_\a{\ga_{\r\b}}^\m\th^{\a\b}-\frac{1}{12}p^\r p_\a {\ga_{\r\b}}^\m\th^{\a\b}-\frac{1}{12}k^\m k^\r {\ga_{\r\a\b}}\th^{\a\b}+\frac{1}{24}p^\m k^\r {\ga_{\r\a\b}}\th^{\a\b}-\frac{1}{24}k^\m p^\r {\ga_{\r\a\b}}\th^{\a\b}\\
&+\frac{1}{12}p^\m p^\r{\ga_{\r\a\b}}\th^{\a\b}+\frac{1}{8}k^2{\ga_{\a\b}}^\m\th^{\a\b}-\frac{1}{8}k\cdot p{\ga_{\a\b}}^\m\th^{\a\b}+\frac{1}{12}k^\r p^\s \ga_{\r\s\b}\th^{\m\b}\Big)P_L,\\
B_9=&-\frac{g^2}{16\pi^2\epsilon}\bigoplus_F\!\Big(\!C_2(F)\!-\!\frac{1}{2}C_2(G)\Big)T^A_F\Big(\frac{1}{6}\ga^\m k_\a p_\b\th^{\a\b}+\frac{1}{6}p^\m k_\a\ga_\b\th^{\a\b}-\frac{1}{4}k^\m p_\a \ga_\b\th^{\a\b}\\
%%%
&+\frac{5}{12}p^\m p_\a \ga_\b\th^{\a\b}+\frac{1}{12}\kslash p_\a\th^{\a\m}+\frac{1}{12}\pslash p_\a\th^{\a\m}-\frac{1}{12}k\cdot p\ga_\b\th^{\m\b}-\frac{1}{12}p^2\ga_\b\th^{\m\b}+\frac{1}{6}p^\r k_\a{\ga_{\r\b}}^\m\th^{\a\b}\\
%%%
&-\frac{1}{4}k^\r p_\a{\ga_{\r\b}}^\m\th^{\a\b}-\frac{1}{12}p^\r p_\a {\ga_{\r\b}}^\m\th^{\a\b}-\frac{1}{8}p^\m k^\r {\ga_{\r\a\b}}\th^{\a\b}-\frac{1}{24}k^\m p^\r {\ga_{\r\a\b}}\th^{\a\b}+\frac{1}{12}p^\m p^\r{\ga_{\r\a\b}}\th^{\a\b}\\
%%%
&+\frac{1}{8}k\cdot p{\ga_{\a\b}}^\m\th^{\a\b}-\frac{1}{12}k^\r p^\s \ga_{\r\s\b}\th^{\m\b}\Big)P_L,
\end{align*}
 In the formulae above, $\mathbb{I}_r$ denotes the identity operator in the linear space defined by the representation $r$.
 
\section{Beta functions of the physical couplings $g$ and $\theta$}	

It is an elementary exercise to work out the one-loop beta functions,
$\beta_g$ and $\beta_\th$, of $g$ and $\theta$, respectively, which are the only physical 
couplings  of the theory. One gets the following results
\begin{align}\nonumber
\beta_g=&-\frac{g^3}{16\pi^2}\Big(\frac{11}{3}c_2(G)-\frac{4}{3}\sum_r c_2(r)\Big),\\
%%%
\label{betas}\beta_\th=&-\frac{g^2\th}{6\pi^2}(C_2(G)-4 C_2(r)).
\end{align}
$C_2(r)$ represents the second-degree Casimir invariant of the representation $r$, while $c_2(r)$ is the index of the representation. Both are related by the relation
\begin{align}\label{Cs}
C_2(r)=c_2(r)\frac{N(G)}{N(r)},
\end{align}
$N(r)$ being the dimension of the representation $r$, and $G$ denoting the adjoint representation.
The $\beta$ function for the gauge coupling $g$ is the same as in the commutative theory. The $\beta$ function for $\theta$, due to the presence of matter, has generically the opposite sign as that of the beta function for the noncommutative parameter that was computed for noncommutative pure gauge theories in ref.~\cite{Latas:2007eu}.  $\beta_\th$ in eq.~\eqref{betas} can be seen to be positive for $E_6$ and $SO(10)$ representations with dimensions less than $100000$ and $12000$, respectively, using the data in ref.~\cite{Slansky:1981yr}.

\end{document}